\documentclass[a4paper]{spie} 
\usepackage[]{graphicx}

\title{The NeXT Mission} 

\normalsize

\author{Tadayuki Takahashi\supit{a}, Richard Kelley\supit{b},
Kazuhisa Mitsuda\supit{a}, Hideyo Kunieda\supit{c}, Robert Petre\supit{b}, Nicholas White\supit{b}, Tadayasu Dotani\supit{a}, 
 Ryuichi Fujimoto\supit{d}, Yasushi Fukazawa\supit{e}, \\ Kiyoshi Hayashida\supit{f},   Manabu Ishida\supit{a},  Yoshitaka Ishisaki\supit{g},  Motohide Kokubun\supit{a}, \\
Kazuo Makishima\supit{h},   Katsuji  Koyama\supit{i}, Greg M. Madejski\supit{j}, Koji Mori\supit{k},  Richard Mushotzky\supit{b}, Kazuhiro Nakazawa\supit{h}, Yasushi Ogasaka\supit{c}, Takaya Ohashi\supit{g},  Masanobu Ozaki\supit{a}, Hiroyasu Tajima\supit{j}, \\
 Makoto Tashiro\supit{l}, Yukikatsu Terada\supit{l}, Hiroshi Tsunemi\supit{f}, 
 Takeshi Go Tsuru\supit{i}, Yoshihiro Ueda\supit{m}, Noriko Yamasaki\supit{a}, Shin Watanabe\supit{a}
and the NeXT team
\skiplinehalf
\supit{a}Institute of Space and Astronautical Science (ISAS), JAXA, Kanagawa, 229-8510, Japan; \\
\supit{b}NASA/Goddard Space Flight Center, Greenbert, MD 20771, USA;\\
\supit{c}Department of Physics, Nagoya University, Nagoya,  338-8570, Japan;\\
\supit{d}Department of Physics, Kanazawa University, Kanazawa,  920-1192, Japan;\\
\supit{e}Department of Physical Science, Hiroshima University, Hiroshia , 739-8526, Japan;\\
\supit{f}Department of Earth and Space Science, Osaka University, Osaka,  560-0043, Japan;\\
\supit{g}Department of Physics, Tokyo Metropolitan University, Tokyo, 192-0397, Japan;\\
\supit{h}Department of Physics, University of Tokyo, Tokyo,  113-0033, Japan;\\
\supit{i}Department of Physics, Kyoto University, Kyoto,  606-8502, Japan;\\
\supit{j}KAVLI Institute, Stanford University, Menlo Park,  CA 94025, USA;\\
\supit{k}Department of Applied Physics, Miyazaki University, Miyazaki, 889-2192, Japan;\\
\supit{l}Department of Physics, Saitama University, Saitama,  338-8570, Japan;\\
\supit{m}Department of Astronomy, Kyoto University, Kyoto,  606-8502, Japan
}

\authorinfo{Send correspondence to T. Takahashi, E-mail: takahasi@astro.isas.jaxa.jp}

\begin{document} 
  \maketitle 

%%%%%%%%%%%%%%%%%%%%%%%%%%%%%%%%%%%%%%%%%%%%%%%%%%%%%%%%%%%%% 
\begin{abstract}

The NeXT (New exploration X-ray Telescope), the new  Japanese X-ray Astronomy Satellite following Suzaku,  is an international
X-ray mission which is currently planed for launch in 2013. 
NeXT is a combination of wide band X-ray spectroscopy (3--80~keV) provided by multi-layer coating, focusing hard X-ray mirrors and hard X-ray imaging detectors, and high energy-resolution soft X-ray spectroscopy (0.3--10~keV) provided by thin-foil X-ray optics and a micro-calorimeter array.  The mission will also
carry an X-ray CCD camera as a focal plane detector for a soft X-ray telescope and a  non-focusing soft gamma-ray detector. With these instruments, NeXT covers very wide energy range from 0.3~keV to 600~keV\@.  The micro-calorimeter system will be developed by international collaboration lead by ISAS/JAXA and NASA\@. The simultaneous broad bandpass, coupled with high spectral resolution of $\Delta E $ $\sim$7~eV 
by the micro-calorimeter will enable a wide variety of important science themes to be pursued.

\end{abstract}

%>>>> Include a list of keywords after the abstract 


%%%%%%%%%%%%%%%%%%%%%%%%%%%%%%%%%%%%%%%%%%%%%%%%%%%\

\section{Introduction}

\begin{figure}
\centerline{\includegraphics[width=12cm,angle=0]{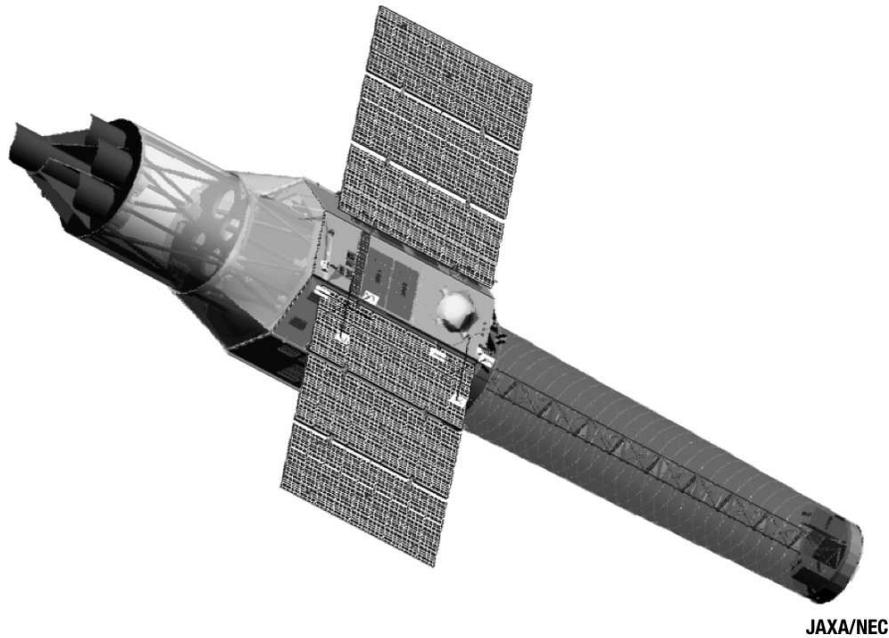}}
\caption{Artist's drawing of the NeXT satellite. The focal  length of the Hard X-ray Telescope (HXT) is 12m, whereas the Soft X-ray Spectrometer (SXS) and Soft X-ray Imager (SXI)
 will have a focal length of 6~meters.}
\label{Fig:NeXT}
\end{figure}

The NeXT (New exploration X-ray Telescope) mission has been studied as the next key X-ray astronomy
mission of Japan, which will be developed under international
collaboration (see Fig.~\ref{Fig:NeXT}).  It has completed the pre-phase
A study in 2006 and entered into Phase A since June 2007.
A Request For Proposal (RFP) for the spacecraft was released by ISAS/JAXA to Japanese
industry in November 2007 and NEC was selected.
 In May, 2008, System Definition Review, which is required before entering Phase B, was completed.
 We  anticipate that NeXT will enter Phase B in the middle of 2008 for the
 planned launch year of 2013. More recently, NASA has selected the US participation in NeXT as a mission of opportunity.   Under this program, the NASA/Goddard Space Flight Center will collaborate with ISAS/JAXA on the implementation of an X-ray microcalorimeter spectrometer\cite{Ref:ProposaNASA}.

The NeXT
mission objectives are to 
study the evolution of yet-unknown obscured super massive
Black Holes (SMBHs) in Active Galactic Nuclei (AGN); trace the growth
history of the largest structures in the Universe;
provide insights into the behavior of material in
extreme gravitational fields; determine the spin
of black holes and the equation of state of neutron
stars; trace particle acceleration structures in
clusters of galaxies and SNRs; and investigate the
detailed physics of jets. 

In this paper, we will summarize the scientific requirements, 
the mission concept and the current
baseline configuration of instruments
of NeXT\@.

\section{Science Requirements}

By comparison with the
soft X-ray band, where the spectacular data from the previous X-ray
 satellites are revolutionizing our understanding of the universe below 10~keV, 
the sensitivities of hard X-ray missions flown so
far  have not dramatically improved
over the last decade. 
The imaging capabilities at high X-ray energies open the completely
new field of spatial studies of non-thermal emission above 10~keV\@. This
will enable us to track  the evolution of Active Galaxies with accretion flows which are
heavily obscured, in order to accurately assess their contribution to
the cosmic X-ray background over cosmic time.
 It will also allow us to
map the spatial extent of the hard X-ray emission in diffuse sources,
tracing the sites of particle acceleration  structures in clusters of Galaxies and
supernova remnants\cite{Ref:Koyama,Ref:Uchiyama,Ref:Aharonian}. 
 Observing the hard X-ray synchrotron  emission
  gives the maximum electron energy produced by the particle
acceleration mechanism in SNR, while the high resolution SXS data on the
gas kinematics of the SNR constrains the energy input into the
accelerator.  

\begin{figure}
\centerline{\includegraphics[width=16cm,angle=0]{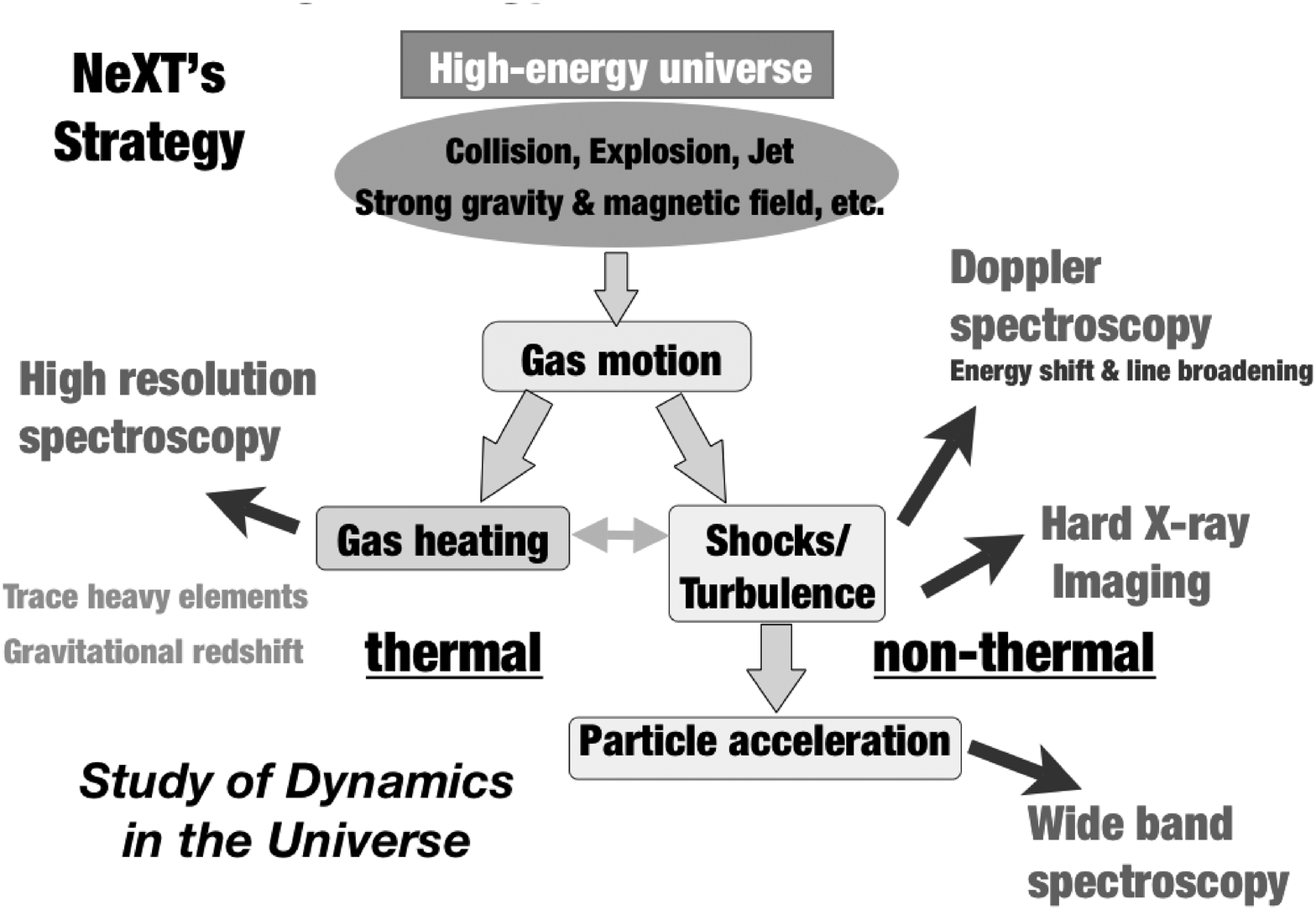}}
\caption{Strategy of the NeXT mission to study high energy universe. A set of detectors, which can perform (1) High resolution spectroscopy, (2) Doppler spectroscopy, (3) Hard X-ray imaging, and
(4) Wide band spetroscopy, are required for the understanding of the high energy universe. }
\label{Fig:Strategy}
\end{figure}

 In order to study the energy content of
non-thermal emission and to draw a more complete picture of the
high energy universe, observations by both  a spectrometer with
an extremely high resolution capable of measuring the bulk plasma velocities and/or turbulence with
the resolution corresponding to the speed of
a few $\times$ 100~km/s and an arc-min imaging system in hard X-ray, 
with the sensitivity two-orders of magnitude better than previous 
missions, are required (See Fig.~\ref{Fig:Strategy}).
In clusters, X-ray hot gas is trapped in the gravitational potential
well  and shocks and/or turbulence are produced in this gas, as smaller
substructures with their own hot gas halo fall into and merge with the
dominant cluster. Large scale shocks can also be produced as gas from the
intracluster medium falls into the gravitational potential of a
cluster.
Here there is  a strong synergy between the hard
X-ray imaging data and the high resolution (several eV) soft X-ray
spectrometer which allows us to study the gas 
kinematics (bulk motion and turbulent velocity)  via the width and energy of 
the emission lines. The kinematics of the gas provides unprecedented
information about the bulk motion; the energy of this motion is in turn
responsible for acceleration of particles to very high energies at
shocks, which is in turn manifested via non-thermal processes, best
studied via sensitive hard X-ray measurement.

On the smallest scales, many active galactic nuclei (AGN) show signatures from the innermost accretion disk in the form of broad ``relativistic" Fe K emission lines. These broad lines were discovered using ASCA in the early 1990s and have been confirmed by XMM-Newton and Suzaku\cite{Ref:YTanaka,Ref:Reeves}. There is, however, a complex relationship between the Fe K line properties, the underlying continuum,
 and the signatures of cold and/or
partially ionized material near the AGN\@. Precise measurements of the complex Fe K line and absorption components require high spectral resolution. Changes
in the X-ray emission and absorption spectral features on the orbital time scale of black holes in AGN, enable  characterization of the velocity field and ionization state of the plasma closest to the event horizon.
The optically thick material that produces the broad fluorescent Fe K line also creates a Compton peak at $E>20$~keV detectable with hard X-ray and soft gamma-ray detectors, providing multiple insights into the physics of the disk. In order to understand the evolution of environments surrounding super massive black holes, 
we need high signal-to-noise measurements of the broad lines of hundreds of AGN up to, at least, $z$$\sim$2. This requires high spectral resolution and bandpass extending to at least $\sim$40~keV\@.  These observations will provide the first unbiased survey of broad Fe K line properties across all AGN\@.

Precision cosmology uses astronomical observations to determine the large-scale structure and content of the Universe. Studies of clusters of galaxies have provided independent measurements of the dark energy equation of state and strong evidence for the existence of dark matter. 
Using a variety of techniques (including the growth of structure, the baryonic fraction in clusters, and the Sunyaev-Zel'dovich effect) a well-constructed survey of clusters of galaxies, with the necessary supporting data\cite{Ref:Rapetti}, can provide precise measurements of cosmological parameters, including the amount and properties of dark energy and dark matter. The key is to connect observables (such as flux and temperature) to cluster mass. In the next few years, cluster surveys will be carried out by 
eROSITA, the South Pole Telescope, the Planck mission, and other Sunyaev-Zel'dovich Effect telescopes. To reduce the systematic uncertainties on the masses inferred from the coarser data from these surveys, a training set of precise cluster masses must be obtained. Measurements of bulk motion of cluster of galaxies and
amounts of non-thermal energies going to cosmic-ray acceleration could reduce the ``{\dots} substantial uncertainties in the baryonic physics which prevents their use at a high level of precision at the present time" (Dark Energy Task Force, Albrecht et al. 2006). Line diagnostics with energy resolution $<$7~eV greatly reduce the uncertainties in the baryonic physics by determining the velocity field, any deviations from thermal equilibrium, and an accurate temperature for each cluster. Information about the non-thermal particle content 
of the cluster can be determined via measurements of their Compton upscattering 
of the CMB:  this is best studied via hard X-ray imaging, providing additional 
clues about the physical state of the cluster gas.

\begin{table}
\caption{NeXT Mission}
\label{Table:Spec}
\begin{small}
\begin{center}
\hspace{5mm}
\begin{tabular}[htp]{|l|l|}
\hline
launch date & 2013 (planned) \\
launch vehicle & JAXA H-IIA \\
orbit & 550 km circular, $<$30 degree\\
\hline
\end{tabular}
\end{center}
\end{small}
\end{table}

\section{Spacecraft and Instruments}

\begin{figure}[htb]
\centerline{\includegraphics[width=17cm]{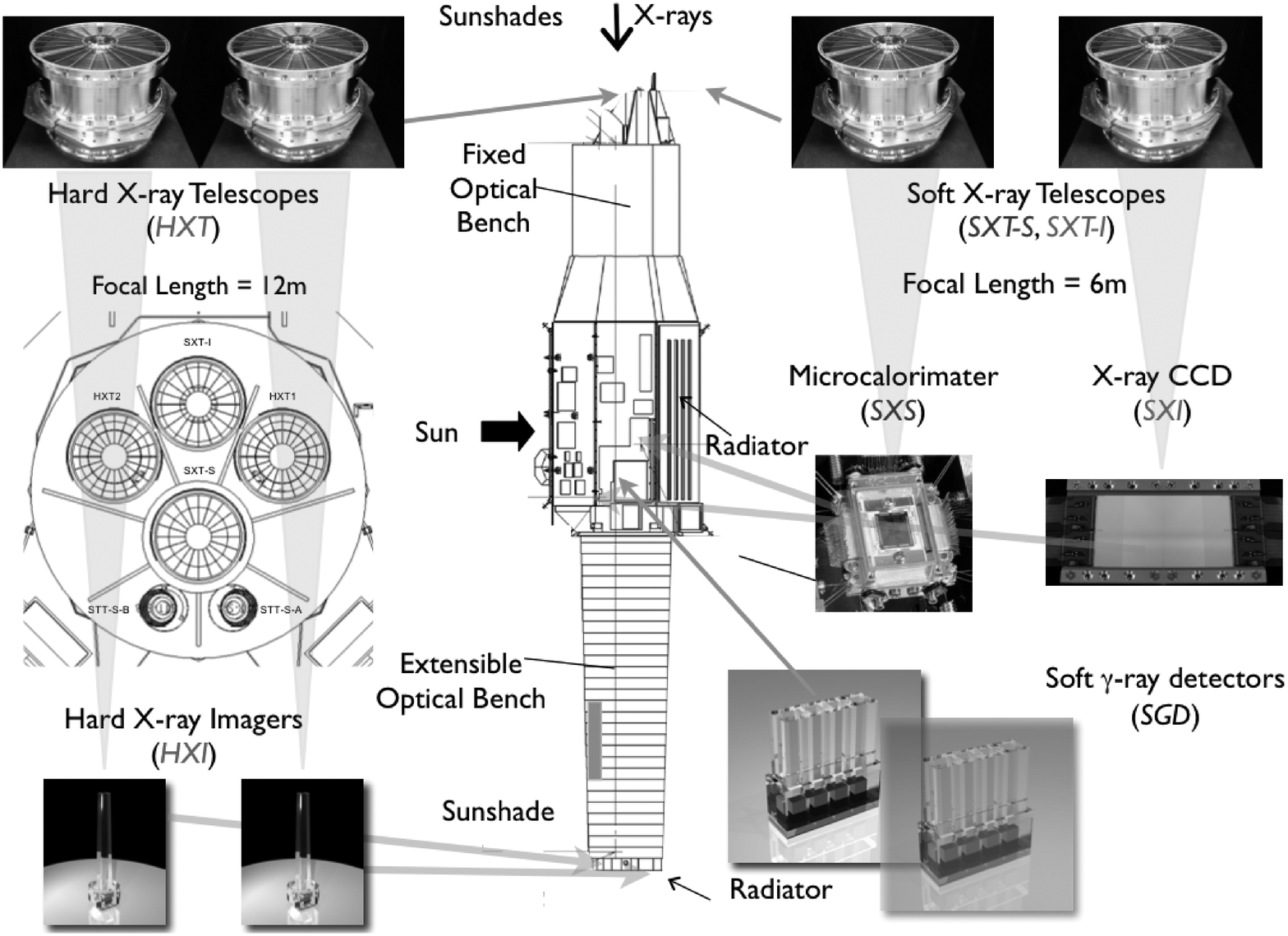}}
\caption{Configuration of the NeXT satellite}
\label{Fig:Config}
\end{figure}

NeXT is designed to perform leading edge science with cutting edge technology.
NeXT will carry
two hard X-ray telescopes (HXTs)
for the hard X-ray imager (HXI), and two soft X-ray telescopes (SXT), one
with a micro-calorimeter spectrometer array  with excellent energy
resolution of $\sim$7\,eV, and the other with a large area CCD\@.
In order to extend the energy coverage to soft $\gamma$-ray region, a Soft
Gamma-ray 
Detector (SGD) will be implemented as a non-focusing detector. 
With these instruments, 
NeXT will cover the bandpass
between 0.5~keV to 600~keV\@.  The
conceptual design of each instrument is shown in Fig.~\ref{Fig:Config}
and the specifications of instruments based on the base-line design are
summarized in Table.~\ref{Table:Spec} and ~\ref{Table:Spec2}.

Both soft and hard X-ray mirrors are mounted on top of the fixed
optical bench. Two focal-plane detectors for soft X-ray 
mirrors are mounted on the base plate of the spacecraft,
while two hard X-ray detectors are mounted on the extensible 
optical bench to attain 12~m focal length.  

NeXT is in many ways similar to Suzaku in terms of orbit,
pointing, and tracking capabilities, although the mass is
 larger; the total mass at launch will be 2400~kg. NeXT will be launched into a
 circular orbit with altitude 500--600~km, and inclination 30~degrees or less. 
 Science operations will be similar to that of Suzaku, with pointed observation
 of each target until the integrated observing time is accumulated, and then
 slewing  to the next target. A typical observation will require 40--100~ksec integrated
 observing time, which corresponds to 1--2.5~days of clock time. All instruments
 operate simultaneously.
 
\begin{table}
\caption{Specification of instruments (Requirements)}
\label{Table:Spec2}
\begin{small}
\begin{center}
\hspace{5mm}
\begin{tabular}[htp]{l|l|ll}
\hline \hline
Hard X-ray Imaging System &\multicolumn{3}{l}{HXT (Hard X-ray Telescope)/HXI (Hard X-ray Imager)} \\ \cline{2-4}
            & &Focal Length & 12 m\\
               & &Effective Area &300 cm$^{2}$ (at 30 keV) \\
      & &Energy Range&5--80 keV \\
              & &Angular Resolution & $<$1.7 arcmin (HPD)  \\
              & &Effective FOV & $\sim$9 $\times$ 9 arcmin (12 m Focal Length)  \\           
 & & Energy Resolution & $<$ 1.5 keV (FWHM, at 60 keV)\\
& & Timing Resolution & several 10 $\mu$s \\
& & Detector Background & $<$ 1--$3 \times 10^{-4}$ cts s$^{-1}$ cm$^{-2}$ keV$^{-1}$ \\
& & Operating Temperature & $<$ $-20$ $^\circ$C \\
\hline \hline
Soft X-ray Spectrometer System&\multicolumn{3}{l}{SXT-S (Soft X-ray Telescope)/SXS (Soft X-ray Spectrometer) }  \\ \cline{2-4}
           & &Focal Length & 6 m\\
  & &Effective Area & 210 cm$^{2}$  (at 6 keV) \\
   & &Energy Range & 0.3--10 keV \\
& &Angular Resolution & $<$ 1.7 arcmin (HPD)  \\
 & &Effective FOV & $\sim$ 3 $\times$ 3 arcmin  \\         
& &Energy Resolution&  $<$7 eV (FWHM, at 7 keV)\\
& & Timing Resolution & several 10 $\mu$s \\
  & & Detector Background &  $<$$5 \times 10^{-3}$ cts s$^{-1}$ cm$^{-2}$ keV$^{-1}$\\
& & Operating Temperature & 50 mK \\
\hline \hline
Soft X-ray Imaging System &\multicolumn{3}{l}{SXT-I (Soft X-ray Telescope)/SXI (Sard X-ray Imager)} \\ \cline{2-4}
           & &Focal Length & 6 m\\
  & &Effective Area & 360 cm$^{2}$  (at 6 keV) \\
               & &Energy Range &  0.3--12 keV  \\
            & &Angular Resolution & $<$ 1.7 arcmin (HPD)  \\
               & &Effective FOV & $\sim$ 35 $\times$ 35 arcmin   \\      
 & &  Energy Resolution & $<$ 150 eV (FWHM, at 6 keV) \\
 & & Timing Resolution & 4 sec \\
  & & Detector Background &  $<$ a few $\times 10^{-3}$ cts s$^{-1}$ cm$^{-2}$ keV$^{-1}$\\
& & Operating Temperature & $-120$ $^\circ$C\\

\hline \hline
Soft Gamma-ray non-Imaging System& \multicolumn{3}{l}{SGD (Soft Gamma-ray Detector)}   \\ \cline{2-4}
    & &Energy Range & 10 keV$-$600 keV  \\
   & &Energy Resolution & 2 keV (FWHM, at 40 keV) \\
   & &Effective Area&  $>$200 cm$^{2}$ Photo absorption mode (at 30 keV)   \\
   & &&  $>$30 cm$^{2}$ (Compton mode, at 100 keV)  \\
   & &FOV  & 0.55 $\times$ 0.55 deg$^{2}$  ($<$ 150 keV) \\
  &  &       & $<$10 $\times$ 10 deg$^{2}$  ($>$ 150 keV) \\
 & & Timing Resolution & several 10 $\mu$s \\
 & & Detector Background  & $<$ a few $\times 10^{-6}$ cts s$^{-1}$ cm$^{-2}$ keV$^{-1}$ \\
 & & & ( $100-200$ keV) \\
& & Operating Temperature & $-20$ $^\circ$C\\
\hline \hline
\end{tabular}
\end{center}
\end{small}
\end{table}

\subsection{Hard X-ray Imaging System}

The hard X-ray imaging system onboard NeXT consists of two identical mirror-detector 
pairs (Hard X-ray Telescope (HXT) and Hard X-ray Imager (HXI)).
The HXT  has conical-foil mirrors with graded multilayer reflecting surfaces 
that provide a 5--80~keV energy range\cite{Ref:Ogasaka}. 
The effective area of the HXT
is maximised for a long focal length, with current design value  of 12~m
giving an effective area of $\sim$300~cm$^{2}$ at 30~keV\@. 
A depth-graded multi-layer mirror reflects X-rays not only by total
external reflection but also by Bragg reflection. In order to obtain
high reflectivity up to 80~keV, the HXT's consist of a stack of
multi-layers with different sets of periodic length and number of layer
pairs with a carbon/platinum coating. The technology of a hard X-ray
focusing mirror has already been proved by the balloon programs
InFOC$\mu$S (2001, 2004)\cite{Ref:Kunieda2006}, 
HEFT (2004)\cite{Ref:Fiona} and SUMIT (2006)\cite{Ref:Kunieda2006}. 

The non-imaging instruments flown so far were essentially limited to
studies of sources with 10--100~keV fluxes of at best $>$10$^{-12}$--%
10$^{-11}$~erg~cm$^{-2}$s$^{-1}$. This limitation is due to the presence
of high un-rejected backgrounds from particle events and Cosmic X-ray
radiation, which increasingly dominate above 10~keV\@. Imaging, and
especially focusing instruments have two tremendous advantages. Firstly,
the volume of the focal plane detector can be made much smaller than for
non-focusing instruments, so reducing the absolute background level
since the background flux generally scales with the size of the
detector.  Secondly, the residual background, often time-variable, can
be measured simultaneously with the source, and can be reliably
subtracted.  For these reasons, a focusing hard X-ray telescope in
conjunction with an appropriate imaging detector sensitive for hard
X-ray photons is the appropriate choice to achive a breakthrough in
sensitivity for the field of high energy astronomy.

The HXI consists of four-layers of 0.5~mm thick Double-sided Silicon Strip
Detectors (DSSD) and one layer of 0.5--1~mm thick CdTe imaging detector (Fig.~\ref{Fig:HXI})\cite{Ref:Nakazawa,Ref:Kokubun}.  
In this configuration, soft X-ray photons will be
absorbed in the Si part (DSSD), while hard X-ray photons go through the
Si part and are detected by the newly developed CdTe double strip detector. With this configuration,
the low energy spectrum, obtained with Si,  is less contaminated with the background due to
activation in heavy material, such as Cd and Te. Also we could use the depth information of 
interactions.  Fast timing
response of silicon strip detector and CdTe strip detector allows us to
place the whole detector inside very deep well of the active shield
made of BGO (Bi$_{4}$Ge$_{3}$O$_{12}$). Signal from the BGO
shield is used to reject background events. The total thickness of the four DSSDs
is 2~mm, the same as that of the PIN detector of the HXD onboard Suzaku.
The DSSDs cover the energy below 30~keV while the 
CdTe strip detector covers the 20-80~keV band.
The DSSD has a size of 3$\times$3~cm$^{2}$
and a thickness of 0.5~mm, resulting in 2~mm in total. A CdTe strip
detector has a size of $\sim$2$\times$2~cm$^{2}$ and a thickness of 0.5~mm. 
In addition to the increase of efficiency, the stack configuration and
individual readout provide information on the interaction depth. This
depth information is very useful to reduce the background in space
applications, because we can expect that low energy X-rays interact in
the upper layers and, therefore, we can reject low energy events
detected in lower layers. Moreover, since the background rate scales
with the detector volume, low energy events collected from the first few
layers in the stacked detector have a high signal to background ratio,
in comparison with events obtained from a monolithic detector with a
thickness equal to the sum of all layers.

  \begin{figure}[htb]
\centerline{\includegraphics[width=17cm]{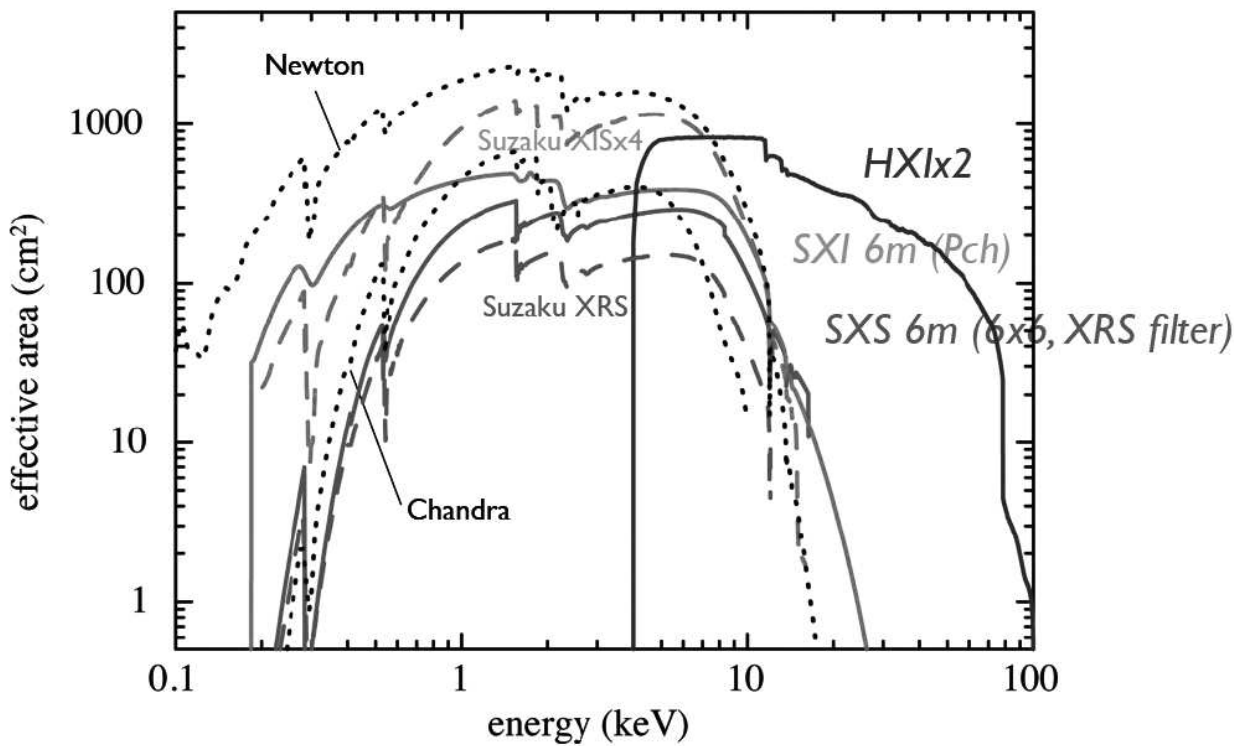}}

\caption{Effective Area of XRT telescopes for the NeXT mission}
\label{fig:area-1}
\end{figure}

\begin{figure}
\centerline{\includegraphics[height=9cm,clip]{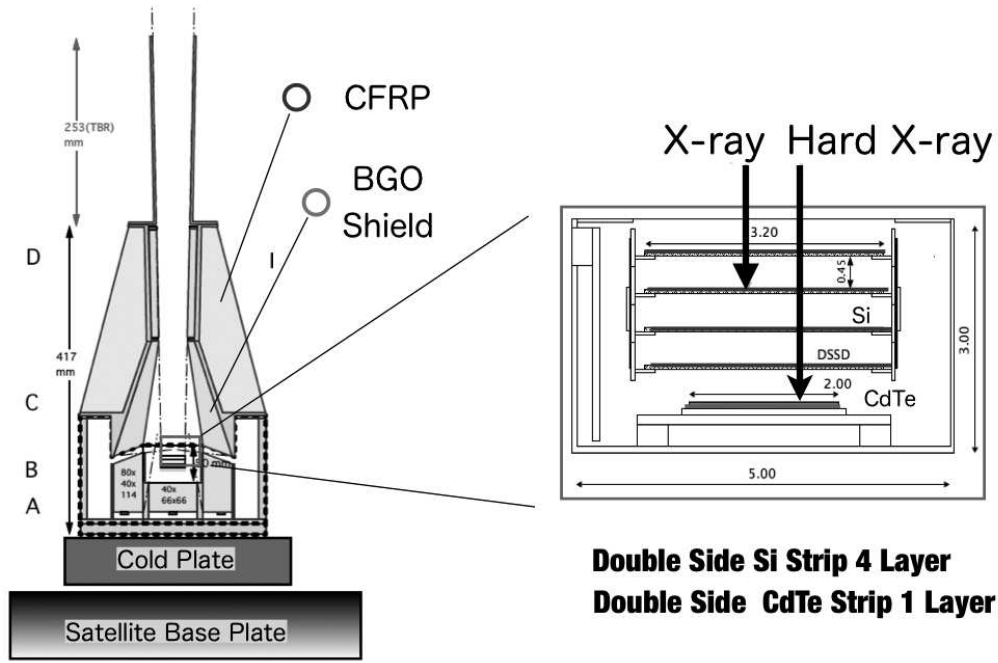}}
\caption{(left) Schematic drawing of the Hard X-ray Imager (HXI) (right) Cross section of the main
detection part of the HXI, placed in a deep BGO active shield. }
\label{Fig:HXI}
\end{figure}

\subsection{Soft X-ray Spectrometer System}

The Soft X-ray Spectrometer  (SXS)  is based on Suzaku-XRS (X-Ray Spectrometer) technology, which is the lowest risk option for implementing the capabilities needed for the SXS\@. The SXS consists of the Soft X-ray Telescope 
(SXT-S), the X-ray Calorimeter Spectrometer (XCS) and the cooling system\cite{Ref:Mitsuda}.

The XCS is a 32~channel system with an energy resolution of $\leq$7~eV between 0.3--12~keV\@. Micromachined, ion-implanted silicon is the basis of the thermistor array, and 8-micron-thick HgTe absorbers provide high quantum efficiency across the 0.3--12~keV band. With a 6-m focal length, the 0.83~mm pixel pitch corresponds to 
0.48~arcmin, giving the array a field of view of 2.85~arcmin on a side. The detector assembly provides electrical, thermal, and mechanical interfaces between the detectors (calorimeter array and anticoincidence particle detector) and the rest of the instrument.

The SXS science objectives require a mirror with larger effective area than those flown on Suzaku,
especially in the Fe K band. This is attained by the combination of increased focal length and larger
diameter. The SXS effective area at 6~keV is 210~cm$^{2}$, a 60~\% increase over the Suzaku XRS,
while at 1~keV the SXS has 160~cm$^{2}$, a 20~\% increase. If we adopt a thin filter, the effective area
at 1~keV increases to $\sim$250~cm$^{2}$. The required angular resolution is 1.7~arcmin, HPD, comparable to the orbit performance of the mirrors on Suzaku.

The XCS cooling system must cool the array to 50~mK with sufficient duty cycle to complete the SXS scientific objectives, requiring extremely low heat loads. To achieve the necessary gain stability and energy resolution, the cooling system must regulate the detector temperature to within 2~$\mu$K rms for at least 24~hours per cycle. As with Suzaku, the array will be cooled using an Adiabatic Demagnetization Refrigerator (ADR). This device is straightforward to construct, has no moving parts or sensitivity to gravity, and has strong flight heritage established by the XRS\@. The ADR and detector assembly will be developed, integrated, and tested together at GSFC, prior to installation into a redundant dewar system developed by ISAS\@. The ADR Controller (ADRC) electronics control and monitor the ADR performance.

For SXS, ISAS/JAXA will provide a redundant dewar system utilizing closed-cycle mechanical coolers that are not limited by expendables. This design is based on coolers developed for space-flight missions in Japan (Suzaku, Akari, and the SMILES instrument to be deployed on the ISS, Narasaki et al. 2005) that have achieved excellent performance with respect to cooling power, efficiency and mass. The 20~K thermal stage consists of a pair of redundant two-stage Stirling-cycle coolers. To provide the sub-2K heat-sink stage for the ADR, there is a hybrid, one-fault-tolerant system consisting of a LHe tank and a 3He Joule-Thomson (JT) refrigerator operating at 1.8~K\@. Under nominal operation, the JT cooler provides a 1.8~K thermal interface and maintains very low heat loads to the He tank. The helium reservoir ensures a high-heat-capacity thermal heat sink for the ADR\@. In the event that the JT cooler fails to operate, the He tank would become the primary heat sink (1.3~K) for the ADR; it is sized to last two years. Conversely, if the LHe supply is rapidly exhausted, the JT cooler would provide adequate cooling for science operations.
A series of five blocking filters shield the calorimeter array from UV and longer wavelength radiation. The aluminized polyimide filters are identical to those successfully flown on Suzaku. 

In combination with a high throughput X-ray telescope, SXS improves on the Chandra and XMM-Newton grating spectrometers in two important ways. At E $>$2~keV, SXS is both more sensitive and has higher resolution (Figs.~1.1 and 1.2), especially in the Fe K band where SXS has 10 times the collecting area and energy resolution, giving a net improvement in sensitivity by a factor of 30 over Chandra. 
The broad bandpass of SXS encompasses the critical inner-shell emission and absorption lines of Fe I-XXVI between 6.4 and 9.1~keV\@. Fe lines are useful because of their (1) strength, due to the high abundance and large fluorescent yield (30\%), (2) spectral isolation from other lines, and (3) relative simplicity of the atomic physics. Fe K emission lines reveal conditions in plasmas with temperatures between 10$^{7}$ and 10$^{8}$ K, typical values for stellar accretion disks, SNRs, clusters of galaxies, and many stellar coronae. In cooler plasmas, Si, S, and Fe fluorescence and recombi-nation occurs when an X-ray source illuminates nearby neutral material. Fe emission lines provide powerful diagnostics of non-equilibrium ioniza-tion due to innershell K-shell transitions from Fe XVII--XXIV\cite{Ref:Decaux}.

SXS uniquely performs high-resolution spectroscopy of extended sources. In contrast to a grating, the spectral resolution of the calorimeter is unaffected by source size because it is non-dispersive. For sources with angular extent larger than 10~arcsec, Chandra MEG resolution is degraded to that of a CCD; that of the XMM-Newton RGS is similarly degraded for sources with angular extent $\geq$2~arcmin. SXS makes possible high-resolution spectroscopy of sources inaccessible to current grating instruments.

\begin{figure}
\centerline{\includegraphics[height=7.5cm,clip]{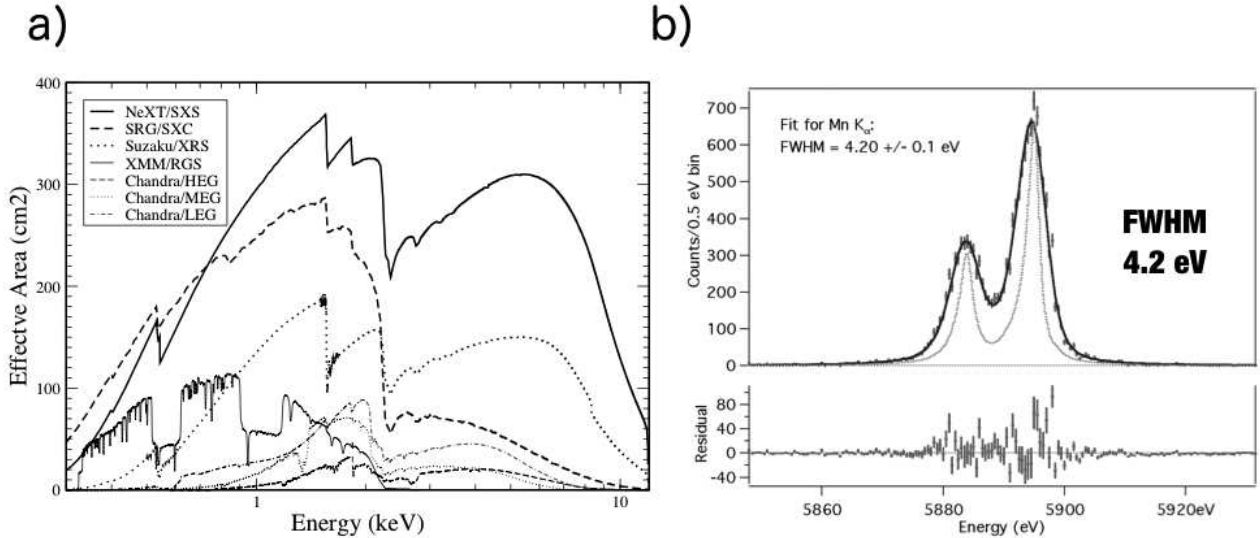}}
\caption{(a) The effective area of SXS (b) The energy resolution obtained from Mn K$\alpha_{1}$using a a detector from the XRS program but with a new sample of absorber material (HgTe) that has lower specific heat, leading to an energy resolution of 4.2~eV (FWHM).  The SXS could have an energy resolution approaching this value (see Kelley et al 2007).}
\label{Fig:XRS}
\end{figure}

The key properties of SXS are its high spectral resolution for both point and diffuse sources over a broad bandpass ($\leq$7~eV FWHM throughout the 0.3--12~keV band), high sensitivity (effective area of 160~cm$^2$ at 1~keV and 210~cm$^2$ at 7~keV), and low non-X-ray background (1.5$\times$10$^{-3}$~cts~s$^{-1}$keV$^{-1}$). These properties open up the full range of plasma diagnostics and kinematic studies of X-ray emitting gas for thousands of targets, both Galactic and extragalactic. SXS improves upon and complements the current generation of X-ray missions, including Chandra, XMM-Newton, Suzaku and Swift. 

\subsection{Soft X-ray Imaging System}
X-ray sensitive silicon charge-coupled devices (CCDs) are a key device
for the X-ray astronomy. The low background and high energy resolution
achieved with the XIS/Suzaku clearly show that the X-ray CCD will also
play very important role in the NeXT mission.  
Soft X-ray imaging system consists of an imaging mirror and a CCD camera
 (Soft X-ray Telescope (SXT-I)  and Soft X-ray Imager (SXI))\cite{Ref:Tsuru,Ref:Tsunemi}.
 
In order to cover the soft
X-ray band below 10~keV,
the SXI will use next generation Hamamatsu CCD chips with
a thick depletion layer, low noise, and almost no cosmetic defects. 
The SXI features a large FOV and covers 35$\times$35~arcmin$^{2}$ region on the sky,
complementing the smaller FOV but much higher spectral resolution of the 
SXS calorimeter. A mechanical
cooler ensures a long operational life at $-$120~$^\circ$C\@. The overall quantum
efficiency and spectral resolution is better than the Suzaku XIS\@.
The imaging mirror has a 6-m focal  length, and a diameter no larger
than 45~cm.

\begin{figure}
\centerline{\includegraphics[height=7.5cm,clip]{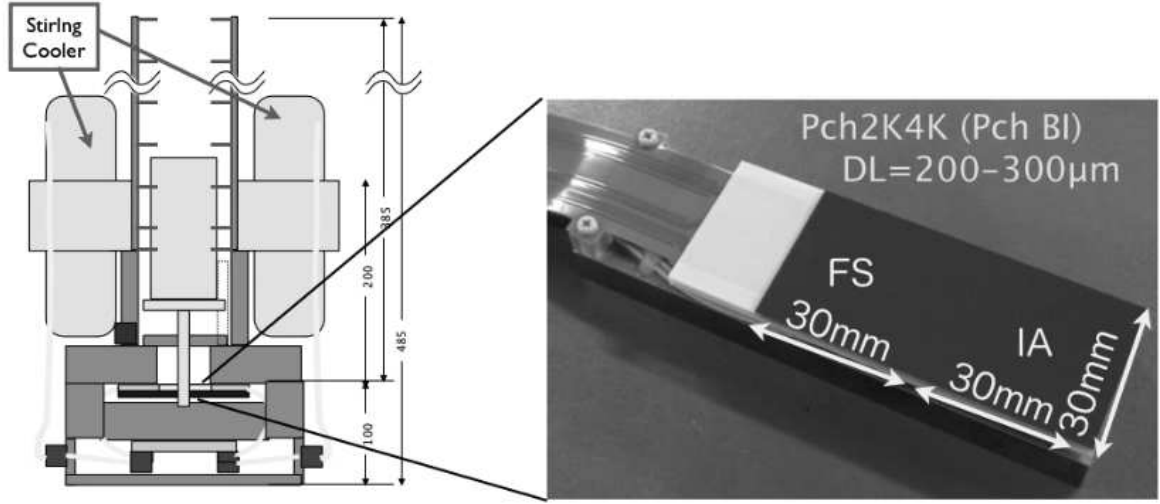}}
\caption{Schematic drawing of the Soft X-ray Imager (SXI and a picture of a prototype CCD chip). }

\end{figure}

\subsection{Soft Gamma-ray Detector (SGD)}

Highly sensitive observations in the energy range above the HXT/WXI
bandpass are crucial to study the spectrum of accelerated particles. 
The SGD is a non-focusing soft gamma-ray detector with a
10--600~keV energy range and sensitivity at 300~keV, more than
10 times better than the Suzaku HXD (Hard X-ray Detector)\@.
It outperforms previous soft-$\gamma$-ray instruments in background
rejection capability by adopting a new concept of narrow-FOV Compton
telescope\cite{Ref:Takahashi_Yokohama,Ref:Takahashi_SPIE2}.

\begin{figure}
\centerline{\includegraphics[height=7.0cm,clip]{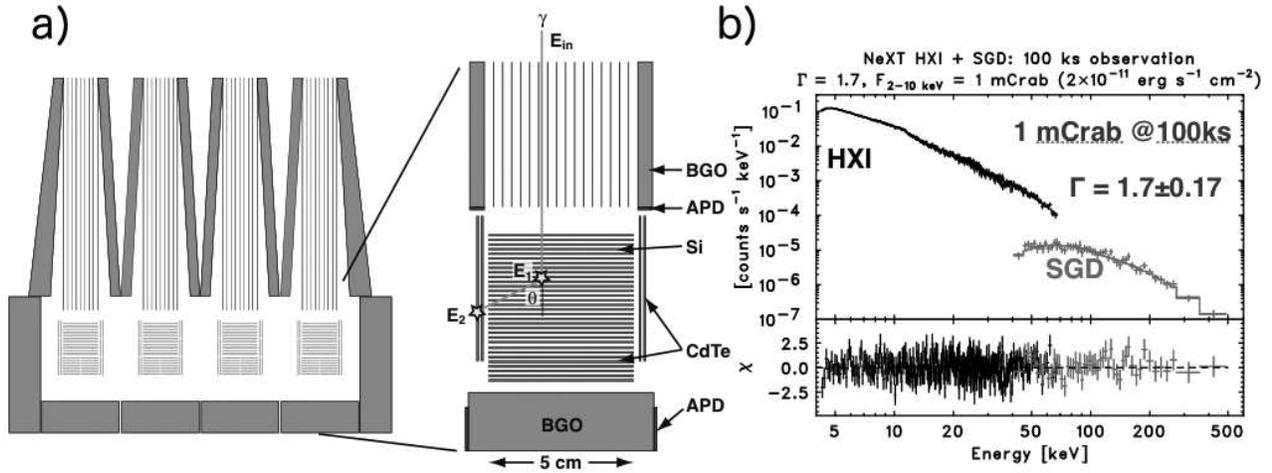}}
\caption{(a) Schematic drawing of the Soft Gamma-ray Detector (SGD). (b) Simulated spectrum of 1~mCrab hard X-ray source for the observation time of 100~ksec.}
\label{Fig:SGD_Concept}
\end{figure}

In order to lower the background dramatically and thus to improve the
sensitivity as compared to the HXD of Suzaku, we combine a stack of Si strip
detectors and CdTe pixel detectors to form a Compton telescope.  The
telescope is then mounted inside the bottom of a well-type active
shield.  Above $\sim$50~keV, we can
require each event to interact twice in the stacked
detector, once by Compton scattering in a stack of Si strip detectors,
and then by photo-absorption in the CdTe part (Compton mode).  Once the
locations and energies of the two interactions are measured, the Compton
kinematics allows us to calculate the energy and direction (as a cone in
the sky) of the incident $\gamma$-ray by following the Compton equation,

As shown schematically in Fig.~\ref{Fig:SGD_Concept} (right), the telescope
consists of 32 layers of 0.6~mm thick Si pad detectors
and eight  layers of thin CdTe pixellated with a thickness of 0.75~mm.  The sides is
also surrounded by two layers of CdTe pixel detectors. The
opening angle provided by the fine collimator is 0.5~degree and 
by the BGO shield is $\sim$10~degrees at 500~keV, respectively. As
compared to the HXD, the shield part is made compact by adopting the
newly developed avalanche photodiode.
An additional copper collimator restricts the field of view of the
telescope to 30' for low energy photons ($<$100~keV) to minimize the
flux due to the Cosmic X-ray Background from the FOV\@.  These modules
are then arrayed to provide the required area (Fig.~\ref{Fig:SGD_Concept} (left)). 
We will have two SGD detectors each consisted of four units. Each
detector will be mounted separately on two sides of the satellite.
It should be noted that when the Compton condition is not used (Photo absorption
mode),
the stacked DSSD  can be used as an usual photo-absorption type
detector with  the total thickness $\sim$20~mm of silicon. The detector then
covers from 10~keV as a collimator-type $\gamma$-ray detector.
The effective area of the SGD is $>$30~cm$^{2}$ at 150~keV in the
Compton mode and 
$>$200~cm$^{2}$ at 30~keV in the Photo absorption mode.
Since the  scattering angle of  gamma-rays can be measured
 when we reconstruct the Compton scattering in the Compton camera, the
SGD will also be sensitive to the polarization of the incident
$\gamma$-rays\cite{Ref:Tajima-pol}.

\begin{figure}[htbp]
\begin{center}
\includegraphics*[width=12cm]{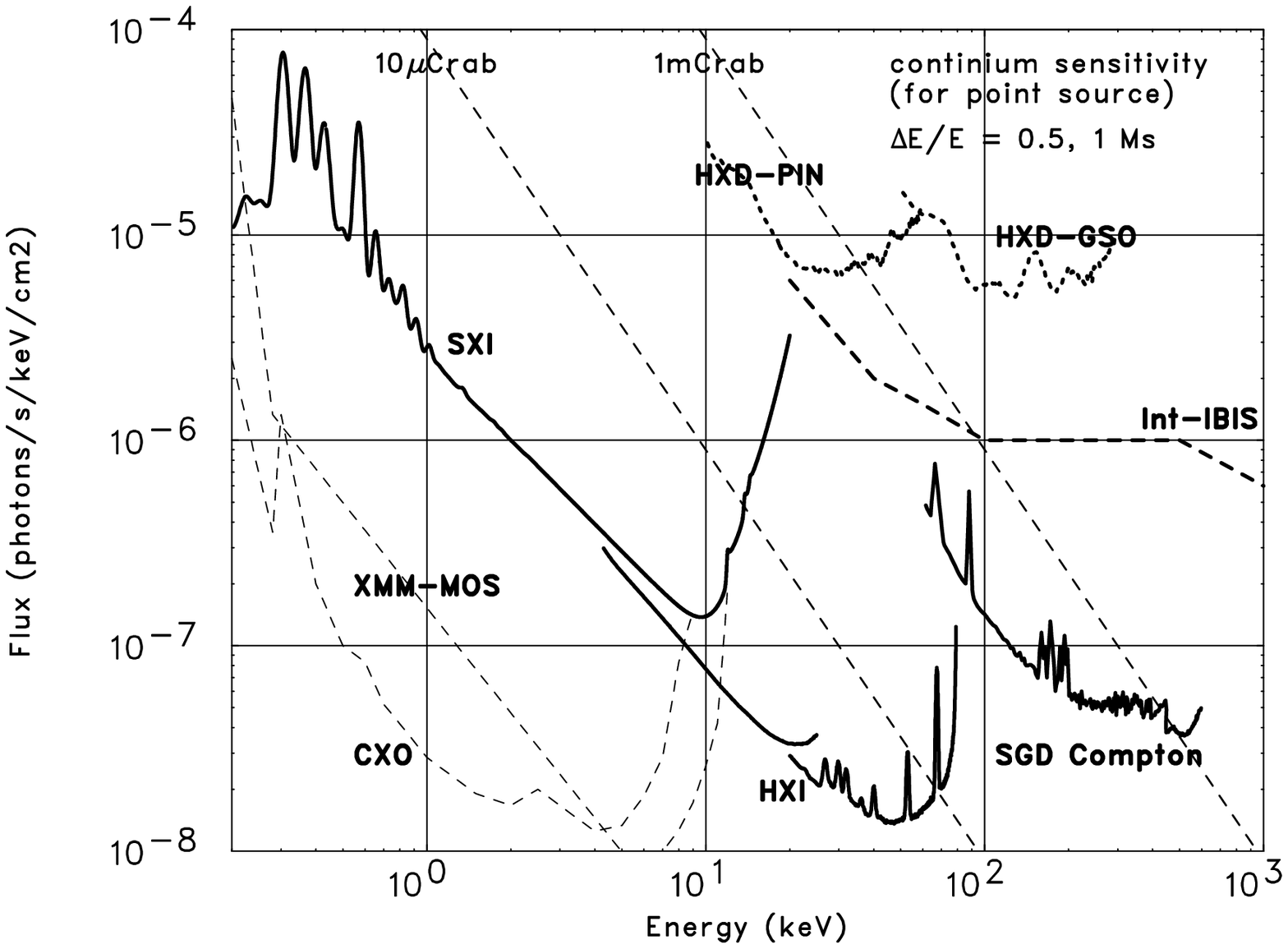}
\vspace{5mm}
\includegraphics*[width=12cm]{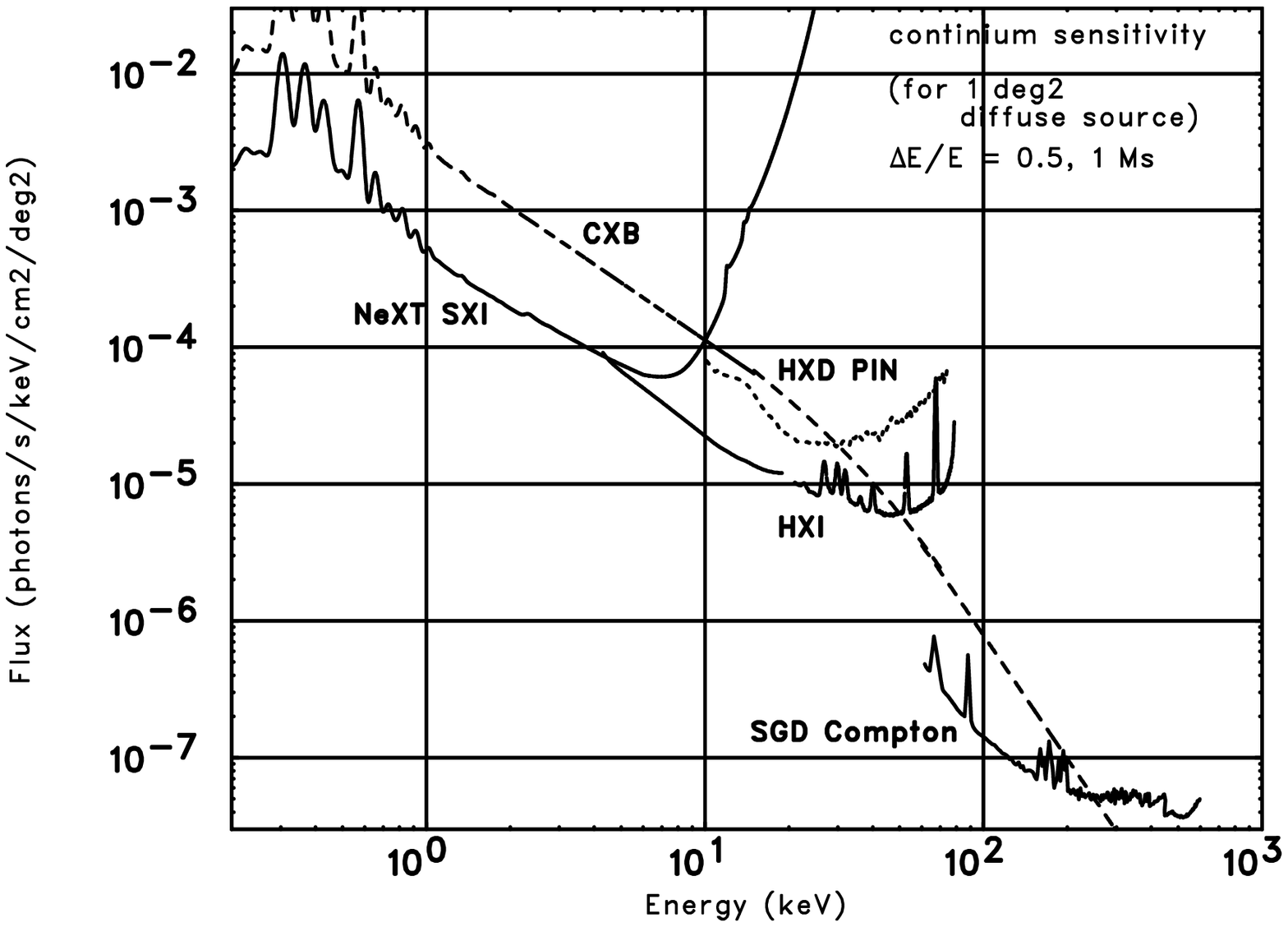}
\end{center}
\caption{Detection limits of the SXT-I/SXI and HXT/HXI for (a) point sources  and (b) for sources with of 1$\times$1~deg$^{2}$
 extension
(bottom) as functions of X-ray energy, where the spectral binning with ${\Delta}E/E = 0.5$ and 1000~ksec exposure are assumed.
Detection limits for the XMM-Newton, Chandra, Suzaku and Integral observatories are shown for comparison for the sensitivity for point sources.}
\label{fig:sensitivity-1}
\end{figure}

\section{Expected Performance}

\begin{figure}
\centerline{\includegraphics[height=7cm,clip]{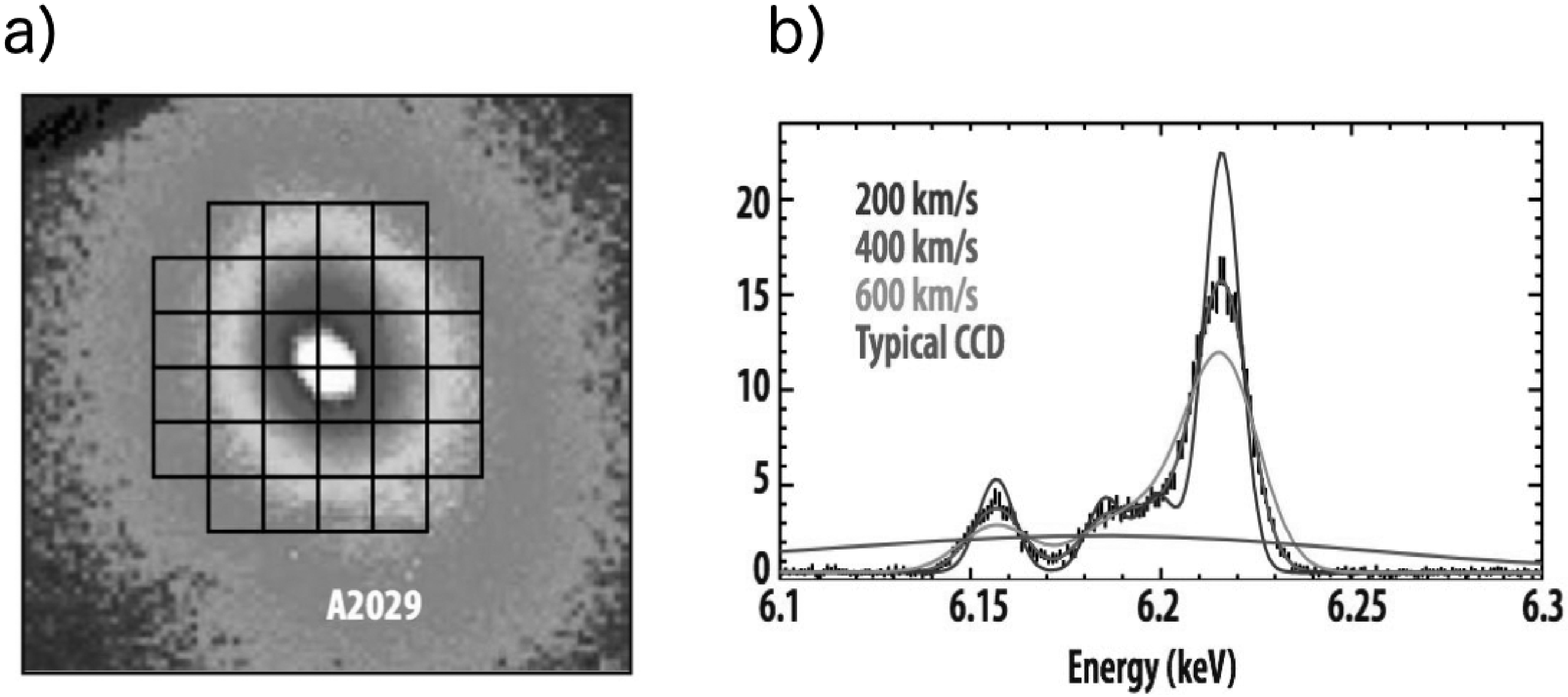}}
\caption{A portion of a simulated spectrum from the cluster A2029,
assuming 400~km/s turbulence, and models assuming 200, 400, and 600
km/s, clearly showing the capability of SXS to measure cluster dynamics. The simulation is for 100~ksec.}
\label{Fig:SXS}
\end{figure}

With NeXT, we expect to achieve an area of about 300\,cm$^{2}$ at
30\,keV with a typical angular resolution of 60$"$ (HPD). Fig.~\ref{fig:sensitivity-1}
shows  Detection limits of the SXT-I/SXI and HXT/HXI for point sources and for
 sources with of 60'  $\times$  60'
 extension.
By assuming a
background level of $\sim$ 1$\times$10$^{-4}$ counts/s/cm$^{2}$/keV, in which
the non X-ray background is dominant, the source detection limit in
1000\,ksec would be roughly 10$^{-14}$\,erg\,cm$^{2}$\,s$^{-1}$ in terms
of the 10--80\,keV flux for a power-law spectrum with a photon index of
2. This is about two orders of magnitude better than present
instrumentation, so will give a breakthrough in our understanding of
hard X-ray spectra. With this sensitivity, 40-50~\% of hard X-ray Cosmic
Background  would be be resolved\cite{Ref:Ueda}.

In addition to the imaging observations below 80~keV, the SGD will
provide a high sensitivity in the soft $\gamma$-ray region to match the
sensitivity of the HXT/HXI combination.  The extremely low background
brought by the new concept of a narrow-FOV Compton telescope adopted for
the SGD will provide sensitive $\gamma$-ray spectra up to 600~keV, with moderate
sensitivity for polarization measurements.

SXS spectroscopy of extended sources can reveal line broadening and Doppler shifts due to turbulent or bulk velocities (Fig.~\ref{Fig:SXS}). This capability enables the spectral identification of cluster mergers, SNR ejecta dispersal patterns, the structure of AGN and starburst winds, and the spatially dependent abundance pattern in clusters and elliptical galaxies. SXS can also measure the optical depths of resonance absorption lines, from which the degree and spatial extent of turbulence can be inferred. Additionally, SXS can reveal the presence of relatively rare elements in SNRs and other sources through its high sensitivity to low equivalent width emission lines. The low SXS background ensures that the observations of almost all line rich objects will be photon limited rather than background limited. 

XMM-Newton and Suzaku spectra frequently show time-variable absorption and emission features in the 5--10~keV band. If these features are due to Fe, they represent gas moving at very high velocities with both red and blue shifted components from material presumably near the event horizon. CCD resolution is too low and the required grating exposures are too long to properly characterize the velocity field and ionization of this gas and determine whether it is from close to the black hole or from high velocity winds. SXS, in combination with  HXI, provides a dramatic increase in sensitivity over Suzaku, enabling measurements that probe the geometry of the central regions of $\sim$50 AGN on the orbital timescale of the Fe producing region 
(for an AGN with a 3$\times$10$^{7}M_{\odot}$ black hole, this is  $\sim$60~$GM_{\odot}/c^{2}$ = 10~ksec).

\section*{Acknowledgement}

The authors are deeply grateful for on-going contributions provided by other members in the NeXT team in Japan and the US:
 L. Angelini,   N.    Anabuki, K. Arnaud,
   H.    Atsukade,
   H.    Awaki,
   A.    Banba,
   M.    Bautz,  G. Brown,  K.-W. Chan, J. Cottam,   J. P. Doty,
   Y.    Ezoe,
   A.    Furusawa,
   F.    Furusawa, M. Galeazzi, K. Gendreau, 
   Y.    Haba, K. Hamaguchi, I. Harrus, 
   T.    Hashimoto,
   J.    Hiraga,
   A. Hornschemeier, U. Hwang, 
   N.    Isobe,
   M.    Itoh, T. Kallman, 
   H.    Katagiri,
   J.    Kataoka,
   M.    Kawaharada,
   N.    Kawai, C. Kilbourne,
   S.    Kitamamo,
   T.    Kohmura,
   M.    Kokubun,
   A.    Kubota,
    J. Lochner, M. Loewenstein, 
   Y.    Maeda,
   H.    Matsumoto,
D. McCammon,   K.    Matsumoto,
   T.    Mihara,
   E.    Miyata,
   T.    Mizuno,
   H.    Mori,
   K.    Mori,
   K. Mukai, 
   H.    Murakami,
   M.    Murakami,
   T.    Murakami,
   Y.    Nakagawa,
   H.    Nakajima,
   M.    Nakajima,
   M.    Nomachi,
   T.    Numasawa, T. Okajima,
   N.    Ohta,
    F. S. Porter, 
   I.    Sakurai,
 R. Sambruna,
   K.    Sato,
   A.    Senda,
   P. Serlemitsos, 
   K.    Shinozaki,
   R. Smith, 
   Y. Soong, 
   H.    Sugita,
   M.    Suzuki,
    A. Szymkowiak
   H.    Takahashi,
   T.    Tamagawa,
   T.    Tamura,
   T.    Tanaka,
   Y.    Tawara,
    Y.    Terashima,
   H.    Tomida,
   Y.    Uchiyama,
   S.    Ueno,
   S.    Uno,
   Y.    Urata,
   K.    Yamaoka,
   H.    Yamashita,
   M.    Yamauchi,
   S.    Yamauchi,
   D.    Yonetoku,
and
   A.    Yoshida.

\end{document}